\documentclass[3p,times,twocolumn]{elsarticle}

\begin{document}

\begin{frontmatter}



\title{ Quark-hadron duality in hydrodynamics: an example $^*$}


\cortext[cor0]{Talk given at 20th International Conference in Quantum Chromodynamics (QCD 17, 3 July - 7 July 2017, Montpellier - FR} \author[label1,label2]{O.V. Teryaev}

\ead{teryaev@theor.jinr.ru}

\address[label1]{ JINR, 141980, Dubna, Russia }

\address[label2]{National Research Nuclear University (NRNU MEPhI) 31, Kashirskoye shosse, 115409 Moscow, Russia}
 \author[label3,label4]{V.I. Zakharov\corref{cor1}}

\fntext[fn1]{Speaker, Corresponding author.}


\ead{vzakharov@itep.ru}

\address[label3]{Institute of Theoretical and Experimental Physics, B. Cheremushkinskaya 25, Moscow}



\address[label4]{School of Biomedicine, Far Eastern Federal University, Vladivostok, 690950, Russia
 }




\pagestyle{myheadings}

\markright{ }

\begin{abstract}

We consider the problem of transferring overall rotation 

of quark-gluon plasma to polarization of hyperons along the 

rotation axis. As a toy theoretical model, we exploit that of pionic superfluidity induced by chemical potentials violating isotopic symmetry. Apparently, the model accounts only for the light degrees of freedom, that is pions. The rotation, however, results in vortices which are infinitely thin in the hydrodynamic approximation. Field theory resolves the singularity and predicts that the core of the vortices is build up on spins of baryons. We review consequences from
the quark-hadron duality in this case. First, an anomalous triangle graph in effective field theory turns to be dual to the vorticity term in the standard hydrodynamic expansion. And, then,
the overall coefficient determining the polarization of baryons is fixed by duality with the triangle graph in the fundamental field theory. \end{abstract}



\end{frontmatter}



\section{Introduction}

Discovery of a non-vanishing polarization of $\Lambda (\bar{\Lambda})$- hyperons \cite{Star07} in heavy-ion collisions is considered to be one of most important experimental observations of recent years. It is quite commonly assumed nowadays that the mechanism behind is production of a rotating fireball of plasma in peripheral collisions, with sunsequent transfer of the rotation to spins of the hyperons. Actually, such a phenomenon was predicted prior to its observation, see \cite{Roga10,Bazn13} and references therein, and more delailed
calculations have been appearing since then, see, e.g., \cite{Beca13,Beca16,Sori17}.

Models considered so far are mostly phenomenological in nature.

There are, however, some general observations that emerge. 

In particular, one can argue that in the ideal-liquid approximation the polarization of hyperons is not generated and one has to
include effects of dissipation \cite{Mont17}. In fact this 

observation seems to be a specific manifestation of a general rule that in the ideal-liquid approximation there is no exchange of
motion between macro- and microscopic degrees of freedom 

\cite{Zakh16}.

In these notes we approach the problem of identifying the mechanism of production of polarized hyperons from
the field-theory side. This has an advantage

of treating fundamental problems,--such as necessity of accounting for dissipation,--in the most transparent way. The price to pay is to rely on a toy field-theoretic model. Indeed, the fundamental QCD
cannot be applied

directly to the problem considerd because of the strong-coupling nature of QCD at large distances. We will utilize a well-known model of pionic superfluidity at a non-vanishing chemical potential $\mu$ violating isotopic symmetry, see \cite{Sons01,Ahar08} and references therein. This model imitates two important properties of QCD plasma, namely, confinement and low viscosity. Thus, the lessons on dynamcs of plasma derived within the model might be relevant to the
realistic case.

One of interesting results is that this model does provide 

us with a definite

mechanism of transfer of the overall rotation to heavy constituents. Moreover the heavy degrees of freedom are not included originally into the model and emerge through dependence on the ultraviolet
cut off of the radiative corrections to the original model. Note that we follow mostly the paper in ref \cite{Tery17}, see also the references therein. In this notes we emphasize novel aspects
 of hadron-quark duality

as applied to hydrodynamics. Commonly, hydrodynamics is viewed as an instrument to account phenomenologically for
the physics in far infrared. We argue that in the case considered hydrodynamic observables are dependent on short distances.
The bridge between hydrodynamics and physics of short distances is provided by vortices which represent singularities in the hydrodynamic
approximation, see also \cite{Sonz04,Metl05,Kiri12,Aris16}.

\section{Hydrodynamics}

  The beauty of hydrodynamics, as of an effective theory,

is its universality.  

The only input is conservation laws and

 expansion in derivatives, i.e. long-wave approximatiion.  

 In most cases, hydrodynamics addresses classical physics.

In case of superfluidity, the two-liquid model bridges quantum physics to the hydrodynamics. The price, in this case, is an introduction of an extra thermodynamic potential.

In a seminal paper \cite{Sons09} it was argued that in fact 

quantum corrections, or loop effects fix some hydrodynamic variables even in absense of superfluidity. Namely, one considers a generic chiral theory, with $U(1)$ chiral anomaly.
In presence of external electromagnetic fields $F_{\alpha\beta}$ the hydrodynamic equations take the form:
\begin{eqnarray} \label{hydro} \partial_{\mu}T^{\mu\nu}~=~eF^{\nu\rho}j^{el}_{\rho}\\ \partial^{\alpha}j_{\alpha}^{el}~=~0\\\label{three} \partial^{\alpha}j_{\alpha}^5~ =~ \frac{\alpha_{el}}{4\pi} \epsilon^{\alpha\beta\gamma\delta} F_{\alpha\beta}F_{\gamma\delta}\\ \partial^{\alpha}s_{\alpha}~\geq~0 \end{eqnarray} where
$s_{\alpha}$ is the entropy current; $T_{\alpha\beta}, j^{el}_{\alpha}, j^5_{\alpha}$  are energy-momentum tensor, electromagnetic current, and
axial current, respectively. All the sources are expanded in derivatives. For example, to zero order in derivatives the currents are expressible in
terms of the corresponding densities, $\rho_{el}, \rho_5$ and 4-velocity $u_{\alpha}$ of an element of the fluid:
\begin{equation} j_{\alpha}^{el}~\approx~\rho_{el}u_{\alpha},~~j^5_{\alpha}~\approx~ \rho_5u_{\alpha} \end{equation} Similarly, the energy-momentum tensor in the zero-order approximation reduces to energy density and pressure.

The crucial finding of Ref. \cite{Sons09} is that 

the condition of growth of entropy, $\partial^{\alpha}s_{\alpha}\geq~0$ cannot be satisfied on the standard expressions for $s_{\alpha}$. The reason is that the anomalous piece in the r.h.s of Eq. (3)
is not positive-definite.

Instead, one has to add both to $s_{\alpha}$ and 

$j_{\alpha}^{el},j_{\alpha}^5$ terms proportional to 

the magnetic field $B_{\alpha}$ and vorticity $\omega_{\alpha}$

where

\begin{equation} B_{\alpha}~=~(1/2)\epsilon_{\alpha\rho\nu\sigma} u^{\rho}F^{\nu\sigma}, ~~ \omega_{\alpha}=(1/2)\epsilon_{\alpha\nu\rho\sigma} u^{\nu}\partial^{\rho}u^{\sigma} \end{equation}
Moreover, the extra terms are uniquely determined within 

the framework described.

The electromagnetic current acquires

a new piece:

\begin{equation}\label{cme}
\big( j_{\alpha}^{el}\big)_{cme}~=~\frac{e^2\mu_5}{2\pi^2}B_{\alpha}~, \end{equation}
where $\mu_5$ is the chemical potential

conjugated with the axial charge.

This is the so called chiral magnetic effect.

The axial current gets an extra term equal to

\begin{equation} \label{cve}
\big(j_{\alpha}^A\big)_{cve}~=~ 
\frac{(\mu^2+\mu_5^2)}{2\pi^2}\omega_{\alpha}~.
\end{equation}

This is the so called chiral vortical effect which will occupy our attention mostly. Note that the overall coefficient, $1/(2\pi^2)$ is given here
for the case of a single (massless) Dirac particle. Algebraically, this coefficient is directly related to the coefficient in front of the anomaly, see the r.h.s. of Eq. (3).

One could wonder, why we are at all interested

in massless particles. Indeed, our aim is to 

find a way of estimating polarization of hyperons which are

massive particles anyhow, and, at first sight hydrodynamics might depend only on these, non-vanshing masses. As we see later,
hydrodynamics does turn sensitive to short distances,

and the polarazation of hyperons woud be finally determined by

physics of (nearly) massless quarks.

\section{From field theory to superfluidity}

\subsection{The toy model}

We will try to get a new insight into dynamics of chiral fluids by considering pion superfluidity induced by isotopic chemical potential, see
\cite{Sons01,Ahar08} and references therein. The chemical 

$\mu, \mu_5$ potentials are introduced 

through adding new pieces to the effective Hamiltonian: 

\begin{equation}\label{model}
\delta H~=~-\mu\bar{q}\gamma_0(\tau_3/2)q-
\mu_5\bar{q}\gamma_0\gamma_5(\tau_3/2)q
\end{equation}

where $q$ are fields of light quarks, $(\tau_3/2)$ is 

the generator of isotopic rotations

around he third axis, $\mu,\mu_5$ are vector and axial-vector chemical potentials, respectevely. We consider small chemical potebtials, $\mu,\mu_5 \ll m_N$. Moreover, for simplicity we consider the limit of exact chiral symmetry, $m_{\pi}~=~0$.

In case of low energies, interactions are

desribed by a simple effective Lagrangian in 

terms of pion fields:

\begin{equation}\label{modelitself}
L_{chiral}~=~\frac{f_{\pi}^2}{4}\Big(D_{\alpha}UD_{\alpha}U^{\dagger}\Big)
\end{equation}

where the matrices $U$ are parametrized in terms of pion fields $\pi^a$: $U~=~\exp\Big(i\pi^a\tau^a/(2f_{\pi})\Big)~~~~(a=1,2,3)$
and the covariant derivatives are given by:

$$D_{\alpha}U~=~\partial_{\alpha}U-i\delta_{\alpha0}
\big(\hat{\mu_L}U-U\hat{\mu}_R\big)$$

Moreover, in the limit of exact chiral symmetry 

the effect of $\mu_V\neq 0$ can be rewritten in terms of

$\mu_A\neq 0$ and vice verse, see, e.g. \cite{Ahar08}. 

For the sake of definiteness we choose $\mu_A\neq 0,\mu_V=0$.

It is then straightforward to see that $ \mu_A$ triggers a negative mode. It is
less trivial that system is stabilized at a non-zero value of the pion condensate, due to  non-linearity of $L_{chiral}$.
 Density of particles is given by

\begin{equation}\label{density}
\rho_5~=~\mu_{5}\cdot f_{\pi}^2~.
\end{equation}

and vanishes in the limit $\mu_5 \to 0$.

There are remarkably simple rules of translation from

ordinary field theory to superfluidity, see, e.g. 

\cite{Lubl10,Kala14} and references therein.

In particular, in field theory there is a contribution of the Goldstone particle to the axial current, $j_{\alpha}^{5,a}~=~f_{\pi}\partial_{\alpha }\pi^a$, where the pion field is virtual. Now, that the
pions are real particles in the ground state, the zeroth component of the current is to coincide with the density (\ref{density}). As a result,
the axial current apparently becomes:

\begin{equation}
j_{\alpha}^5~=~\rho_5u_{\alpha}~,
\end{equation}

where for the 4-velocity $u_{\alpha}$ we immediately have

\begin{equation}\label{pionhydro}
\mu_5u_{\alpha}~=~\partial_{\alpha}\big(\pi^0/f_{\pi}\big)~.
\end{equation}

For the spatial components, ($\alpha=1,2,3$), 

Eq. (\ref{pionhydro}) becomes a standard expression

for the 3-velocity of supefluid since $\pi^0/f_{\pi}$

is nothing else but the phase of the superfluid  wave function, see the general expression for the matrices $U$ above.

As far as the spatial components $u_i$ are small,

 we have for the phase:

\begin{equation}
\varphi(x_i,t)~=~\frac{\pi^0}{f_{\pi}}~\approx~\mu_5\cdot t+
\tilde{\varphi}(x_i)~,
\end{equation}

where $\tilde{\varphi}(x_i)$ satisfies $\Delta\tilde{\varphi}(x_i)=0$.

Moreover, the $T_{0i}$ components of the

energy-momentum tensor for the field $\pi^0$ are

given by $T_{0i}~\sim~\mu_5\partial_i\tilde{\varphi}$.

And for the correlator of components of the energy-momentum tensor in the momentum space one gets:
\begin{equation}\label{criterion}
\lim_{q_i\to ~0,\omega\equiv~ 0}{<T_{0i},T_{0k}>}~
\sim~\mu_5^2\frac{q_iq_k}{q_i^2}~.
\end{equation}

Eq (\ref{criterion}) is nothing else but the criterion of superfluidity.

\section{Duality between chiral vortical effect and spin of vortices}

Consider now rotating superfluid. As is well known, naively, rotation cannot be transferred to supefuid since $\mu_5\vec{v}~=~\vec{\nabla}\varphi$ and $curl\vec{v}=0$. However, rotation is still possible because of vortices, or singularities on the axis of a vortex. Near the singularity,
\begin{equation}\label{vicinity}
\frac{\pi^0}{f_{\pi}}~=~\mu_5\cdot t+\kappa \theta~,
\end{equation}

where $\theta$ is the polar angle and $\kappa$ is integer. Everywhere, with exception of the axis, $curl\vec{v}=0$. However,
\begin{equation}
(\partial_x\partial_y-\partial_y\partial_x)\theta~=~2\pi\delta(x,y).
\end{equation}

The fact that $\kappa$ is integer follows from the condition that the phase of the wave function is to be a single-valued function. In this way one comes to the quantization condition:
\begin{equation}\label{quantization}
\mu_5\oint v_idx^i~=~2\pi \mathcal{\kappa}~, ~~\mathcal{\kappa}= 1,2... \end{equation}
It might worth mentioning that in most applications $\kappa=1$,

as it follows from minimization of energy.

Note that the size of the 

core of the vortex can be sent to zero and still Eq. (\ref{quantization}) holds. Thus, one can say that condition (\ref{quantization}) fixes spin of vortices.

It is worth emhasizing that the spin of vortices is {\it not} the whole angular momentum carried by the superfluid. Indeed, the angular momentum carried by the fluid in the vicinity of  the singularity is given by:
\begin{equation}\label{angular}
\mathcal{L}~=~\int_0^{d_{vortex}}(mass)\cdot r\cdot v_{\phi}~,
\end{equation}

where $d_{vortex}$ is a cut off. The physical meaning of this

cut off is the distance to the next vortex. One can readily see

that the angular momentum (\ref{angular}) is proportional to

the area per vortex. Indeed:

\begin{equation}
\mathcal{L}~\sim~\int_0^{d_{vortex}}(\mu_52\pi rdr)v_{\phi}(r)r~\sim~d^2_{vortex} \end{equation}
This angular momentum is much larger than the spin of the vortex (\ref{quantization}). Moreover, it is not quantized at all and one can say that it is rather classical than quantum.

To appreciate this point, we need to be more specific about the set up considered. Following textbooks, see, e.g., \cite{Land59,Nozi99} we consider a cylindrical bucket with superfluid
rotated around its axis with angular velocity $\omega_z$. 

Then at very low $\omega_z$ the rotation is not transferred to the fluid at all. With increase of $\omega_z$ vortices get excited. There exists a range of values of $\omega_z$ when the distance between the vortices is still much larger than the size of their cores, on one hand, and, on the other hand, the total angular momentum carried by
the supefluid is (very) close to

the value of the angular momentum which would be carried by a solid body of the same form and mass.

The number of vortices, $n_{vortex}$ is determined by the condition of matching the motions of superfluid
and of the analogous solid body.

Then for the number of vortices penetrating area $A$

in the $(x,y)$ plane one gets \cite{Kiri12}:

\begin{equation}\label{number}
n_{vortex}~=~\frac{\mu_5}{\pi}\int_Ad^2x|\omega_z|~, 
\end{equation} 

Averaging over vortices locally allows then to introduce a

quasi-continuum picture. As the next step, we would like to evaluate the total spin of vortices and compare it
with the chiral vortical effect (\ref{cve}) found above

within the standard hydrodynamic approach to

continuum, or defect-free media.

To this end let us explain how one can derive

Eq. (\ref{cve}) in field theory.

First, we adjust the model (\ref{model}) for the use within hydrodynamic approach. This adjustment is achieved, as usual with the help
of the 4-vector $u_{\alpha}$. Namely, one can say that 

the model (\ref{model}) corresponds to the following 

term in the density of the

effective Lagrangian:

\begin{equation}\label{model1}
\delta L~=~\mu u_{\alpha}\bar{q}\gamma^{\alpha}(\tau_3/2)q~+~\mu_5 u_{\alpha}\bar{q}\gamma^{\alpha}\gamma_5(\tau_3/2)q ~.
\end{equation}

A crucial point is the apparent similarity of this extra term, specific for hydrodynamics, with the standard electromagnetic interaction,
$\delta L_{el}~=~eA_{\alpha}\bar{\Psi}\gamma^{\alpha}\Psi$.

This similarity can be accounted for \cite{Sado11} 

by extending the external field 

$eA_{\alpha}$ to the following combination:

\begin{equation}
eA_{\alpha}~\to~eA_{\alpha}+\mu u_{\alpha}~~.
\end{equation} 

If, for example, from field theory we know the contribution

of the anomalous triangle graph with electromagnetic external fields, then we can read off from it
the answer for the graphs with $\mu u_{\alpha}$ as external fields.

As a result we get the following extra term in the axial current evaluated in the hydrodynamic approximation: \begin{equation}\label{extraterm}
\delta j^5_{\alpha}~=~\frac{1}{4\pi^2f_{\pi}^2}
\epsilon_{\alpha\beta\gamma\delta}
(\partial^{\beta}\pi^0)(\partial^{\gamma}\pi^0\partial^{\delta}\pi^0)~.
\end{equation}

Moreover, the field $\pi^0$ in the vicinity of the axis of a vortex is given by Eq. (\ref{vicinity}). As a result,
the axial current induced by a single vortex is

given by:

\begin{equation}\label{singlevortex}
\delta j^5_z~=~\frac{\mu_5\kappa}{2\pi}\delta(x,y)~,
\end{equation}

where in fact $\kappa = 1$ in case considered (see discussion above).

Finally,  combining the equation for the contribution of

a single vortex, see (\ref{singlevortex}), and Eq. (\ref{number}) for the number of vortices one finds \cite{Kiri12}
\begin{equation}\label{match}
(spin~of~vortices)~=~\frac{\mu^2}{2\pi^2}\int d^3x |\vec{\omega}|~. \end{equation}
Remarkably enough, this expression fits perfect the expression for the chiral vortical effect (\ref{cve}), as is observed first in Ref. \cite{Kiri12}. 

Thus, we have established an example of a 

dual description of the same phenomenon in hydrodynamics and field theory. Namely, Eq. (\ref{cve}) can be obtained by evaluating a triangle ``anomalous'' graph with external field $\mu_5\cdot u_{\alpha}$ replacing the standard electromagnetic filed, $eA_{\alpha}$. This evaluation assumes the medium be continuous and is usually thought of as a next term in
expansion in derivatives, i.e. in the

long-wave approximation. Within 

such a framework, however, it is not clear, why

a next order in hydrodynamic expansion would contain an extra Planck constant (corresponding to a loop graph). In this section we have evaluated spin of defects which corresponds to singular fluid configurations and found out that the chiral vortical effect fits the value of spin carried by the defects, or vortices. The presence of the extra power of the Planck constant is crystal clear in this picture, since the quantization condition (\ref{quantization}) does contain it explicitly.

\section{Chiral effects as radiative corrections}

In the hydrodynamic approximation, the vortices are infinitely thin. On the other hand, as we have just discussed, the same
effects can be evaluated within the effective field theory, 

and one might hope that the field theory allows to 

resolve the problem of the ultraviolet cut off.

The main input from field theory is provided by Eq. (\ref{pionhydro}) which relates pion field to the hydrodynamic current. Interactions of the pions are well understood, and therefore there should be no difficulties of pronciple to evaluate radiative corrections to the hydrodynamic superfluid current.

Interactions of pions among themselves are encoded, of course, in the model (\ref{modelitself}) which we are using. However, it can readily be seen that the model in fact is not closed against the radiative corrections. While the very phenomenon of the superfluidity can be established by considering the light degrees of freedom alone, evaluation
of the radiative corrections asks for inclusion of interactions with heavy particles, or baryons.

Consider first the chiral magnetic effect, see Eq. (\ref{cme}). It is convenient to start from the Goldstone-Wilczek current \cite{Gold81,Call85}:
  \begin{equation}\label{gw}
j_{\alpha}^{el}~=~(const)e^2 \epsilon_{\alpha\beta\gamma\delta}
(\partial^{\beta}\pi^0)F^{\gamma\delta} ~.
\end{equation}

Substituting $\partial_0\pi^0~=~\mu_5$ we immediately 

get the electromagnetic current (\ref{gw}) in the hydrodynamic form which fits Eq. (\ref{cme}).

In the original model (\ref{modelitself}) accounting for the

light degrees of freedom, both the electromagnetic current and 

axial current satisfy isotopic selection rules $\Delta I~=~1$

and current (\ref{gw}) cannot be generated. One needs

electromagnetic transitions with both $\Delta I= 0$ and $\Delta I=1$. The $\Delta I =0$ transitions
can be realized only on heavy particles, or baryons.

Therefore, we need to include interaction of pions with baryons.

Within the effective field theory the interaction of the 

Goldstone particles is given by

\begin{equation}\label{hadronic}
L_{effective}~=~f_{\pi}(\partial_{\alpha}\pi^a)\tilde{J}^{\alpha}_{a,heavy},~~~a=1,2,3
\end{equation}

where $\tilde{J}^{\alpha}_{a,heavy}$ is the axial current of heavy particles which is not
conserved kinematically,

 as a result of spontaneous symmetry breaking.

Hence,  baryons

 propagate inside

loops. It is straightforward to check that the loop graphs

converge on virtual momenta of order $p~\sim~1/m_N$. 

Thus, although the original model (\ref{modelitself}) does not include heavy particles, theory of chiral effects requires account of
baryons which emerge

as an ultraviolet cut off.

We are getting also new insight into the role of chiral anomalies in theory of the chiral effects. In Ref. \cite{Sons09} the anomaly introduced into the r.h.s. of the hydrodynamic equations, see (\ref{hydro}), as an input to evaluate the coefficients in the hydrodynamic expansion.
Now we see that by evaluating the chiral effects as radiative corrections within field theory we come to consider exactly the same graphs that determine the $\pi^0\to 2\gamma$ decay and which are very well known from numerous studies of the anomalies.
In a way, the use of

the general hydrodynamic expansion (\ref{hydro}) is cut short.

It s worth emphasizing that implicitly 

such an  approach has been utilized in literature. 

In particular, the famous chiral effective Lagrangian of Witten \cite{Witt81} incorporates the effect of graphs describing the $\pi^0\to 2\gamma$ decay. And, then, this effective Lagrangian was adjusted to the
 hydrodynamic approach in a number of papers,

see, in particular, \cite{Metl05,Lubl10,Kala14}.

  Turn now to the

chiral vortical effect. It is given by the matrix element of the

axial baryonic current ${j}^{\alpha}_{5,baryons}$ with $\Delta I=0$, evaluated in one-loop approximation.
Up to the overall coefficient, which we discuss later, 

the loop graph reduces to the same expression (\ref{extraterm}). In the preceding section we proceeded from Eq. (\ref{extraterm}) to evaluation of spin of vortices. Now, within field theory, we emphasize that the spatial components of the axial current represent spin of baryons. Then we come to an estimate:
\begin{equation}\label{central}
<\vec{(\sigma)}_{baryons}>~\sim~\frac{\mu_5^2}
{2\pi^2}\vec{\omega} \int d^3x
\end{equation}

where $\vec{\omega}$ is the angular velocity of the overall rotation of plasma.

Eq. (\ref{central}) is one of central results of paper \cite{Tery17}. One can say that this result reflects duality between hydrodynamic and effective
field-theory approaches to chiral effects. What has not been fixed yet, is the overall coefficient in
Eq. (\ref{central}). This coefficient is fixed by the hadron-quark duality. To appreciate the point, turn again to the $\pi^0\to 2\gamma$ decay. In the effective theory (\ref{hadronic}) the corresponding
graphs are convergent, and, 

at frst sight, the overall coefficient depends on how many species of baryons we keep in  the
current $\tilde{J}^a_{heavy}$ entering

the Eq. (\ref{hadronic}). But in fact, the overall coefficient is fixed by the quark triangle
graph, or by the hadron-quark duality.

And this coefficient controls the chiral magnetic effect,

see (\ref{cme}).

In case of the chiral vortical effect, the overall coefficient 

(as is  determined by the quark triangle)

is found, e.g., in Ref. \cite{Kala14}:

\begin{equation}
<j^{\alpha}_{5,baryons}>~=~\frac{N_c}{36\pi^2f_{\pi}^2}
\epsilon^{\alpha\beta\gamma\delta}(\partial_{\beta}\pi^0)
(\partial_{\gamma}\partial_{\delta}\pi^0) ~,
\end{equation} 

where $N_c$ is the number of colors. We can then normalize the spin of baryons to the density of particles, $\rho_5$:
\begin{equation}
<\sigma_z>_{per~particle}~\sim~\omega_z\frac{\mu_5}{6\pi^2f_{\pi}^2}~.
\end{equation}

It is worth emphasizing, however, that the medium itself consists of (pseudo)scalar particles, pions, while the spinning component of the plasma is due to baryons.

\section{General theorems vs model calculations}

We are now in position to compare the results obtained within the model (\ref{modelitself}) with general theorems mentioned
in the Introduction.

Turn first to the chiral magnetic effect (\ref{cme}) as derived

in Ref. \cite{Sons09}. We do reproduce this result within the model (\ref{modelitself}) considered. However, the logic of derivation is very diferent. Namely, we start with the tree-level  conserved axial current,
in the hydrodynamic incarnation:

\begin{equation}
\big( J_{\alpha}^5\big)_{hydro}~=~f_{\pi}\partial_{\alpha}\pi^0~\equiv ~\rho_5u_{\alpha}~,
\end{equation}

where $\rho_5$ is defined in Eq. (\ref{density}).

According to the standard current algebra, divergence of the axial current is igiven by
$\partial^{\alpha}J_{\alpha}^5~=~f_{\pi}m_{\pi}^2\pi^0$ and

vanishes for massless pions,

also off-mass shell. This is the so called Sutherland-Veltman theorem \cite{Suth67,Velt67,Dolg67}. The chiral anomaly, on the other hand, implies that even for massless pions
\begin{equation}\label{anomalous}
\partial^{\alpha}J_{\alpha}^5=
f_{\pi}m_{\pi}^2\pi^0+(\alpha_{el}/4\pi)F_{\beta\gamma}\tilde{F}^{\beta\gamma}~.
\end{equation}

The second, anomalous piece can be interpreted as resulting from a direct coupling of $\pi^0$ to electromagnetic fields, $(const)\pi^0F_{\alpha\beta}\tilde{F}^{\alpha\beta}$. Eq (\ref{anomalous}) is used as an input in hydrodynamic equation (\ref{three}). The Goldstone-Wilczek current (\ref{gw}) responsible for the chiral magnetic effect can be obtained by varying  with respect to the electromagnetic
potential of the direct coupling of the $\pi^0$-meson to photons, $(cons)\pi^0F\tilde{F}$.
Thus, what is brought new by the model (\ref{modelitself}) is the factorization of large and short distances. Infrared physics is responsible for the hydrodynamic tree-level current, while short distances determine coupling of the $\pi^0$-field to the electromagnetic potentials. Within this model, connection between the chiral anomaly (\ref{anomalous}) and the chiral magnetic effect (\ref{cme}) becomes a kind of obvious since both effcts are directly detremined by one and the same coupling of $\pi^0$ to electromagnetic fields. To the contrary, the derivation of the chiral vortical effect assumes existence of defects which is an extra dynamical feature.

The difference in derivation of the CME and CVE effects becomes even more striking if we try to include effect of non-vanishing viscosity. In case of the chiral magnetic effect the coupling f $\pi^0$ to electromagnetic fields aparently does not depend on viscocity and one expects that this effect survives also with account of the viscosity. On the other hand, low viscocity is crucial for existence of vortices and, therefore, there is no reason to expect that the chiral vortical effect survives with account of the viscocity. 

Another point to mention, in the Introduction we quoted Refs. \cite{Mont17, Zakh16} according to which transfer of overall rotation to spin of constituents is not allowed without account for dissipation.
The mechanism of such a transfer discussed in these notes does satisfy this general constraint.
Indeed, one of the common and general 

ways to introduce dissipation in the field-theoretic

language is to include interaction of light degrees of freedom, which dominate dynamics of the fluid, with heavy degrees of freedom, see, e.g., \cite{Endl13}. Moreover, interaction of the light degrees of freedom with heavy ones is described by polynomials. The mechanism of generating polarizarion of the hyperons discussed above belongs just to this kind of models. Therefore, the dissipation is indeed playing a crucial role.

 \subsection{Acknowledgments}

We are thankful to A. Avdoshkin, T. Kalaydzhyan, V.P.~Kirilin, A.V.~Sadofyev, J.~Sonnenschein and A.S.~Sorin for interesting discussions.
The work was supported by Russian Science

Foundation Grant No 16-12-10059.


\begin{thebibliography}{999}






\bibitem{Star07}	

STAR Collaboration (B.I. Abelev et al.), Phys.Rev. C76 (2007) 024915, Erratum: Phys.Rev. C95 (2017), 039906,
arXiv:0705.1691 [nucl-ex].

\bibitem{Roga10}

O. Rogachevsky, A. Sorin, and O. Teryaev, Phys. Rev. C82 (2010) 054910, arXiv:1006.1331 [hep-ph] .
\bibitem{Bazn13}

M. Baznat, K. Gudima, A. Sorin, and O. Teryaev, Phys. Rev. C88 (2013), 061901, arXiv:1301.7003 [nucl-th].
\bibitem{Beca13} 	 

F. Becattini, L. Csernai, and D.J. Wang, Phys. Rev. C88 (2013), 034905, Erratum: Phys.Rev. C93 (2016), 069901,
  arXiv:1304.4427 [nucl-th].



\bibitem{Beca16}

F. Becattini, I. Karpenko, M. Lisa, I. Upsal, and S. Voloshin, arXiv:1610.02506 [nucl-th] .


 \bibitem{Sori17} 

arXiv:1606.08398 [nucl-th] .



\bibitem{Mont17}

D. Montenegro, L. Tinti, and G. Torrieri,

``{\it The ideal relativistic fluid limit for a medium with polarization }'', arXiv:1701.08263.


\bibitem{Zakh16}

 V.I. Zakharov, 	

``{\it Notes on conservation laws in chiral hydrodynamics}'',

  arXiv:1611.09113.



\bibitem{Sons01}

D.T. Son and M. A. Stephanov, Phys. Rev. Lett. { 86} (2001) 592,

 hep-ph/0005225.

 

\bibitem{Ahar08}

O. Aharony, K. Peeters, J. Sonnenschein, and M. Zamaklar, JHEP 0802 (2008) 071, arXiv:0709.3948 [hep-th].


\bibitem{Tery17}

O. V. Teryaev and V. I. Zakharov,

{\it ``Chiral vortical effect in pionic superfluid vs spin alignment of baryons''},
e-Print: arXiv:1705.01650. 



\bibitem{Sonz04}

D.T. Son and A. R. Zhitnitsky, Phys. Rev. D{70} (2004) 074018, hep-ph/0405216.


\bibitem{Metl05}

M. A. Metlitski and A. R. Zhitnitsky, Phys. Rev. D72 (2005) 045011, hep-ph/0505072.


\bibitem{Kiri12}

V.P. Kirilin, A.V. Sadofyev, and V.I. Zakharov, Phys. Rev. D86 (2012) 025021, arXiv:1203.6312 [hep-th]




\bibitem{Aris16}

A. Aristova, D. Frenklakh, A. Gorsky, and D. Kharzeev,  

 JHEP 1610 (2016) 029, 

arXiv:1606.05882 [hep-ph] .



\bibitem{Sons09}

D.T. Son and P. Surowka,	 

Phys. Rev. Lett. 103 (2009) 191601, 

arXiv:0906.5044 [hep-th] .



\bibitem{Lubl10}

M. Lublinsky and   I. Zahed, Phys.Lett. B684 (2010)  119,

 arXiv:0910.1373 [hep-th].

 

\bibitem{Kala14}

T. Kalaydzhyan, Phys.Rev. D89 (2014) 105012, arXiv:1403.1256 [hep-th].


\bibitem{Land59}

L. D. Landau and E. M. Lifshitz, {\it Statistical Physics}, Part 2, Pergamon, New York (1959).


\bibitem{Nozi99}

 Ph. Nozieres, D. Pines, {\it Superfluid Bose Liquids},

Theory Of Quantum Liquids, vol. II , Advanced Books Classics, (1999).


\bibitem{Sado11}

A.V. Sadofyev, V.I. Shevchenko, and V.I. Zakharov, 	

Phys. Rev. D83 (2011) 105025, 

arXiv:1012.1958 [hep-th].



\bibitem{Gold81} 	

J. Goldstone and F. Wilczek,

 Phys. Rev. Lett. 47 (1981) 986.



\bibitem{Call85}

C. G. Callan and  J. A. Harvey,

 Nucl. Phys. B250 (1985) 427.



\bibitem{Witt81}

E. Witten, Nucl.Phys.  B223 (1983) 422.



\bibitem{Suth67}

Nucl. Phys. B2, (1967) 433.





\bibitem{Velt67}

M. Veltman, 

Proc. Roy. Soc. A301 (1967) 107.





\bibitem{Dolg67}

A.D. Dolgov, A.I. Vainshtein , and V.I. Zakharov,  

Phys. Lett. 24B (1967) 425.



\bibitem{Endl13}	

S. Endlich, A. Nicolis, R. A. Porto, and J. Wang, Phys. Rev. D88 (2013) 105001, arXiv:1211.6461 [hep-th].




 











 




























 













\end{thebibliography}
\end{document}